\documentclass[reprint,aps,prd,twocolumn,showpacs,superscriptaddress,groupedaddress,amsmath,amssymb,]{revtex4}

\usepackage{graphicx}
\usepackage{dcolumn}
\usepackage{bm}
\usepackage{hyperref}
\usepackage{verbatim}

\newcommand{\ee}[1]{\!\times\!10^{#1}}
\newcommand{\gws}{gravitational-waves }
\newcommand{\gw}{gravitational-wave }

\newcommand{\beq}{\begin{equation}}
\newcommand{\eeq}{\end{equation}}
\newcommand{\nbs}{narrow-band search }
\newcommand{\nbss}{narrow-band searches }

\def\be{\begin{equation}}
\def\ee{\end{equation}}

\begin{document}


\title{A method for narrow-band searches of continuous gravitational wave signals}

\author{P. Astone$^{1}$,  A. Colla$^{2,1}$, S. D'Antonio$^{3}$, S. Frasca$^{2,1}$, C. Palomba$^{1}$ and R. Serafinelli$^2$}
\affiliation{INFN, Sezione di Roma, P.le A. Moro, 2, I-00185 Rome, Italy\\$^2$Dip. di Fisica, Universita' di Roma "Sapienza", P.le A. Moro, 2, I-00185 Rome, Italy\\$^3$INFN,
Sezione di Roma 2}






\date{\today}

\begin{abstract}
Targeted searches of continuous waves from spinning neutron stars normally assume that the frequency of the gravitational wave signal is at a given known ratio with respect to the rotational frequency of the source, e.g. twice for an asymmetric neutron star rotating around a principal axis of inertia.
In fact this assumption may well be invalid if, for instance, the gravitational wave signal is due to a solid core rotating at a slightly different rate with respect to the star crust.
In this paper we present a method for {\it narrow-band} searches of continuous gravitational wave signals from known pulsars in the data of interferometric detectors. This method assumes source position is known to high accuracy, while a small frequency and spin-down range around the electromagnetic-inferred values is explored. Barycentric  and spin-down corrections are done with an efficient time-domain procedure. Sensitivity and computational efficiency estimates are given and results of tests done using simulated data are also discussed.  
\end{abstract}



\maketitle

\section{\label{sec:intro}Introduction}
Continuous gravitational wave signals (CW) emitted by an asymmetric rotating neutron stars are among the sources currently searched in the data of interferometric gravitational wave detectors. Various mechanisms have been proposed that could allow for a time varying mass quadrupole in these stars, thus producing CW.
Typically, CW searches are divided in {\it targeted}, when the source position and phase parameters are known with high accuracy, like in the case of known pulsars, and {\it blind} in which those parameters are unknown and a wide portion of the parameter space is explored. In fact, also intermediate cases can be considered, see e.g. \cite{ref:palMor} for a review of recent results. While {\it targeted} searches can be done using coherent methods, based on matched filtering or its variations, {\it blind} searches are usually performed with hierarchical approaches which strongly reduce the needed computing power at the cost of a relatively small sensitivity loss. 

Targeted searches typically rely on accurate measures of pulsar parameters, among which the rotational frequency and its time variation (spin-down), that come from electromagnetic observations, like those done by radio-telescopes. This means that a strict correlation between the gravitational wave frequency and the measured star rotational frequency is assumed. In the classical case of a non-axisymmetric neutron star rotating around one of its principal axes of inertia the gravitational frequency would be exactly twice the rotation frequency of the star. In fact, that such strict correlation holds for observation times of months to years is questionable and various mechanisms could break this assumption.

In this paper we present a coherent search method that relaxes this assumption allowing for a small mismatch, a fraction of Hertz wide, between the gravitational frequency and two times the rotational frequency (and similarly for the spin-down parameters). For this reason such kind of search is called narrow-band. Until now \nbss have not received much attention, one notable exception being the Crab pulsar search done over LIGO S5 data \cite{ref:crab_nb}. The analysis method we will discuss is based on a computationally efficient way to perform barycentric (Doppler and relativistic effects) and spin-down corrections, first devised in \cite{ref:livas}, followed by a re-sampling of the data, and on matched filtering in the space of signal Fourier components. Such techniques have been already employed for targeted searches \cite{ref:vela_vsr2}, \cite{ref:aasi2013} but their extension and application to narrow-band searches is presented here for the first time. Conceptually, the same method we use for Doppler correction has been described in \cite{ref:pink} where, however, it is implemented in a different way and is used in the context of a different analysis procedure. Another similar method for barycentric corrections, but using data at full bandwidth, has been presented in \cite{ref:stefano}.
 
The plan of the paper is the following. 
In Sec.\ref{sec:signal} we remind the main characteristics of CW. The next three sections of the paper are devoted to describe the main steps of the analysis pipeline used for the targeted search of CW from known neutron stars, of which the narrow-band search method is an extension. In Sec. \ref{sec:bary} an efficient procedure to make barycentric and spin-down correction is described in detail. In Sec. \ref{sec:targeted} the {\it 5-vectors} method, based on matched filtering in the space of signal Fourier components is briefly reminded. In Sec. \ref{sec:pvalue} the way of assessing detection significance is discussed. Following sections are dedicated to present the narrow-band search pipeline. In Sec.\ref{sec:motiv} we explain the motivations for narrow-band searches. In Sec.\ref{sec:anagen} we describe in detail the narrow-band search method. In Sec. \ref{sec:sensi} the narrow-band search sensitivity is computed. In Sec. \ref{sec:test} the validation tests done using simulated data are discussed. Finally, conclusions and future prospects are presented in Sec.\ref{sec:concl}. 

\section{\label{sec:signal}Continuous gravitational wave signals from spinning neutron stars}

The expected quadrupolar \gw signal at the detector from a non-axisymmetric neutron star steadily spinning about one of its principal axis is 
at twice the rotation frequency $f_{rot}$, with a strain of \cite{ref:fivevect}
\be
h(t)=H_0(H_+A^++H_{\times}A^{\times})e^{\jmath \left(\omega_0(t)t+\Phi_0\right)}
\label{eq:hoft}
\ee
where taking the real part is understood. The signal frequency and phase at time $t_0$ are, respectively, $f_0=\frac{\omega_0(t_0)}{2\pi}=2f_{rot}(t_0)$ and $\Phi_0$. The two complex amplitudes $H_+$ and $H_{\times}$ are given respectively by
\begin{equation}
H_+=\frac{\cos{2\psi}-\jmath \eta \sin{2\psi}}{\sqrt{1+\eta^2}}
\label{eq:Hp}
\end{equation}
\begin{equation}
H_{\times}=\frac{\sin{2\psi}+\jmath \eta \cos{2\psi}}{\sqrt{1+\eta^2}}
\label{eq:Hc}
\end{equation}
in which $\eta$ is the ratio of the polarization ellipse semi-minor to semi-major axis and the polarization angle $\psi$ defines the direction of the major axis with respect to the celestial parallel of the source (measured counterclockwise). The parameter $\eta$ varies in the range $[-1,1]$, where $\eta=0$ for a linearly polarized wave and $\eta=\pm 1$ for a circularly polarized wave ($\eta=1$ if the circular rotation is counterclockwise). 
The functions $A^+$ and $A^{\times}$ describe  the detector response as a function of time and are given by
\begin{align}
A^+ =  & a_0+a_{1c} \cos{\Omega_{\oplus} t}+a_{1s} \sin{\Omega_{\oplus} t} +a_{2c} \cos{2\Omega_{\oplus} t}+\\ \nonumber
& a_{2s} \sin{2\Omega_{\oplus} t}  
\label{eq:Ap}
\end{align}
\begin{align}
A^{\times} =  &b_{1c} \cos{\Omega_{\oplus} t}+b_{1s} \sin{\Omega_{\oplus} t} +b_{2c} \cos{2\Omega_{\oplus} t}+\\ \nonumber 
& b_{2s} \sin{2\Omega_{\oplus} t}
\label{eq:Am}
\end{align}
where $\Omega_{\oplus}$ is the Earth sidereal angular frequency and with the coefficients depending on the source position and detector position and orientation on the Earth \cite{ref:fivevect}. 
As discussed in \cite{ref:vela_vsr2} the strain described by Eq.(\ref{eq:hoft}) is equivalent to the standard expression, see e.g. \cite{ref:jks}:
\begin{align}
h(t) =  &\frac{1}{2}F_+(t, \psi)h_0(1+\cos{}^2\iota)\cos{\Phi(t)} \\ \nonumber 
& + F_{\times}(t, \psi)h_0\cos{\iota}\sin{\Phi(t)}
\label{eq:hclass}
\end{align}
depending on the ``classical'' beam-pattern functions $F_+,~F_{\times}$, on the amplitude 
\begin{equation}\label{eq:h0}
h_0=\frac{4\pi^2G}{c^4}\frac{I_{zz}\varepsilon f^2_0}{d}
\end{equation}
in which $I_{zz}$ is the star moment of inertia with respect to the  
principal axis aligned with the rotation axis and $\varepsilon=\frac{I_{xx}-I_{yy}}{I_{zz}}$ is the equatorial ellipticity expressed in terms of principal moments of inertia,
and on the angle $\iota$ between the star rotation axis and the line of sight, given the following relations:
\begin{equation}
H_0=h_0 \sqrt{\frac{1+6 \cos^2 \iota+ \cos^4 \iota}{4}}
\end{equation} 
\begin{equation}
\eta=-\frac{2 \cos \iota}{1+\cos^2 \iota} 
\end{equation}

In fact, in  Eq.(\ref{eq:hoft}) the signal angular frequency $\omega_0(t)$ is a function of time,  and then the signal phase 
\begin{equation}
\Phi(t)=\int_{t_0}^t \omega_0(t')dt'
\label{eq:phit}
\end{equation} 
is not that of a simple monochromatic signal and depends on both the intrinsic rotational frequency and frequency derivatives of 
the pulsar and on Doppler and propagation effects. These effects include relativistic modulations 
caused by the Earth's orbital and rotational motion\footnote{For a source in a binary system also the binary orbital motion must be taken into account.} and the presence of massive bodies in the solar system close 
to the line-of-sight to the pulsar. As a consequence the power of a monochromatic \gw signal would be spread across a range of frequencies, thereby reducing the signal detectability. Hence, these effects must be removed, as described in Sec.(\ref{sec:bary}), before computing a detection statistic. 

Equating the \gw  luminosity 
\begin{equation}
\dot{E}_{gw}=\frac{32\pi^6}{5}\frac{G}{c^5}f^6_0I^2_{zz}\epsilon^2
\label{dotEgw}
\end{equation}
to the kinetic energy lost as the pulsar spins-down ($\dot{E} = 4\pi^2 I_{zz} 
f_{\rm rot}|\dot{f}_{\rm rot}|$, where $\dot{f}_{\rm rot}$ is the star's rotational frequency derivative) 
gives us the so-called {\it spin-down limit} on \gw strain
\begin{align}\label{eq:h0sd}
h_0^{\rm sd} = &\left(\frac{5}{2} \frac{G I_{zz} \dot{f}_{\rm rot}}{c^3 d^2 f_{\rm rot}} 
\right)^{1/2}\nonumber \\
 = & 8.06\times 10^{-19}\frac{I^{1/2}_{38}}{d_{\rm kpc}}\sqrt{\frac{|\dot{f}_{\rm rot}|}{f_{\rm rot}}},
\end{align}
where $I_{38}$ is the star's moment of inertia in the units of $10^{38}$\,kg\,m$^2$, and $d_{\rm kpc}$ is the 
distance to the pulsar in kiloparsecs. The spin-down limit on the signal amplitude corresponds (via 
Eq.~\ref{eq:h0}) to an upper limit on the star's fiducial ellipticity
\begin{equation}\label{eq:eps_sd}
\varepsilon^{\rm sd} = 0.237\left(\frac{h_0^{\rm sd}}{10^{-24}}\right)f_{\rm rot}^{-2} I_{38}^{-1} d_{\rm 
kpc}.
\end{equation}
This quantity, for a given neutron star equation of state, can be put in relation to the physical ellipticity of the star surface \cite{ref:macda}. On the other hand, the $l=m=2$ mass quadrupole moment $Q_{22}$ is related to the \gw amplitude through  \cite[see e.g.][]{ref:owen}
\begin{equation}\label{eq:q22}
Q_{22} = \sqrt{\frac{15}{8\pi}} I_{zz}\varepsilon = h_0\left( \frac{c^4 d}{16\pi^2G f_{\rm rot}^2} \right) 
\sqrt{\frac{15}{8\pi}}.
\end{equation}
This value can be constrained independently of any assumptions about the star's equation of state and its 
moment of inertia.

Setting a \gw   upper limit below the spin-down limit is an important achievement as it allows us to constrain the fraction of spin-down energy due 
to the emission of \gws, which gives insight into the spin-down energy budget.

\section{\label{sec:bary}Barycentric and spin-down corrections}

As anticipated in previous section, the signal frequency at the detector is modified by various effects, the most important of which is the Doppler effect.
The received frequency $f(t)$ is related to the emitted frequency $f_0(t)$ by the well-known relation (valid in the non-relativistic approximation)
 \begin{equation}
f(t)=\frac{1}{2\pi}\frac{d\Phi(t)}{dt}= f_0(t) \left(1+\frac{\vec{v}
\cdot \hat{n}}{c}\right),
\label{eq:fdopp}
\end{equation}
where $\vec{v}=\vec{v}_{rev}+\vec{v}_{rot}$ is the detector velocity with respect to the Solar system barycenter (SSB), sum of the Earth revolution velocity around the Sun, $\vec{v}_{rev}$, and of the Earth rotation velocity, $\vec{v}_{rot}$, while $\hat{n}$ is the versor identifying the source position and $c$ is the light velocity.   
From Eq.(\ref{eq:fdopp}) we see that the frequency variation due to the Doppler effect depends on the frequency itself. This means that, in principle, if the signal frequency is not accurately known in advance and a range of frequencies must be explored, for each given search frequency we have a different correction to compute in order to remove the Doppler effect. 

In practice it is much more efficient to compute the Doppler correction in the time domain. Let us assume that the emitted signal is monochromatic with frequency $f_0$, that is we neglect spin-down for the moment. By integrating Eq.(\ref{eq:fdopp}), and using Eq.(\ref{eq:phit}), we have
\begin{align}
\Phi(t)=2\pi\int_{t_0}^t f_0\left(1+\frac{\vec{v(t')}\cdot \hat{n}}{c}\right)dt'= \\ \nonumber
\Phi_0+2\pi f_0\left(t+\frac{\vec{r(t)}\cdot \hat{n}}{c}\right)    
\label{eq:Phidop}
\end{align}
where $\vec{r}$ is identifies the detector position with respect to the SSB and the initial signal phase is $\Phi_0=-2\pi f_0\left(t_0+\frac{\vec{r(t_0)}\cdot \hat{n}}{c}\right)$.
From the previous equation we immediately see that if we introduce a new time variable 
\begin{equation}
\tau_1=t+\frac{\vec{r(t)}\cdot \hat{n}}{c}=t+\Delta_R
\label{eq:tau1}
\end{equation}
the signal phase, expressed in terms of $\tau_1$, is that of a monochromatic signal:
\begin{equation}
\Phi(\tau_1)= \Phi_0+2\pi f_0 \tau_1
\label{eq:phitau1}
\end{equation}
The correction term $\Delta_R$ is the well-known {\it Romer delay}, which amounts up to about $1,000$ seconds over one year, corresponding to the time taken by a signal traveling at the speed of light to cover the distance between the detector and the barycenter of the solar system. The key point that makes the use of a re-scaled time preferable for Doppler correction in narrow-band searches is that Eq.(\ref{eq:tau1}) {\it does not depend on the frequency}. This means that one single correction holds for every frequency.  
In fact there are other smaller relativistic effects that must be taken into account when making barycentric corrections. One is the {\it Einstein delay},  $\Delta_E$, which takes into account the time delay of special relativity due to the Earth motion and the gravitational redshift at the Earth geocenter due to all the solar system bodies (except the Earth). The Einstein delay is given by 
\begin{equation}
\Delta_E\simeq \frac{1}{c^2}\int_{t_0}^t\left(U_{\oplus} +\frac{v^2_{\oplus}}{2}\right)dt'
\label{eq:eindel}
\end{equation}
where $U_{\oplus}$ is the gravitational potential at the geocenter, due to all solar system bodies, except the Earth, and $v_{\oplus}$ is the velocity of the geocenter with respect to the SSB.  
This integral cannot be computed analytically. In practice we use an approximate series expansion where only the main contributions are considered \cite{ref:irwfuk}.
The Einstein delay amounts to about 2 milliseconds at most. Another effect is the {\it Shapiro delay} $\Delta_S$ which takes into account the deflection of a signal passing near a massive bodies. The main contribution in the solar system comes from the Sun for which we have
\begin{equation}
\Delta_S=-\frac{2GM_{\odot}}{c^3}log{\left(1+cos\theta\right)}
\label{eq:shadel}
\end{equation}
being $\theta$ the angle between the Sun-source direction and the Sun-detector direction. In fact this effect can be shown to negligible for CW searches, unless the source line of sight passes very near the Sun limb, in which case a delay up to about 120 $\mu s$ can be accumulated. 
Overall, we can make the full barycentric corrections by introducing the re-scaled time
\begin{equation}
\tau_1=t+\Delta_R+\Delta_E-\Delta_S
\label{eq:baricorr}
\end{equation}
This transformation corresponds to referring the data collected at the detector site at the SSB, which can be considered an inertial reference frame to a very good approximation.
   
We can take into account the spin-down in a similar way. The frequency evolution due to spin-down can be written as
\begin{equation}
f(t)=f_0+\dot{f}_0(t-t_0)+\frac{1}{2}\ddot{f}_0(t-t_0)^2
\label{sd}
\end{equation}
where higher order terms have been neglected.
The corresponding phase evolution is given by
\begin{align}
\Phi_{sd}(t)=2\pi \int_{t_0}^t f(t')dt'=\\ \nonumber 
\Phi_{sd,0}+2\pi \left(f_0(t-t_0)+\frac{1}{2}\dot{f}_0(t-t_0)^2+\frac{1}{6}\ddot{f}_0(t-t_0)^3\right)
\label{eq:Phisd}
\end{align}
By re-scaling time according to
\begin{equation}
\tau_2=t+\frac{\dot{f}_0}{2f_0}(t-t_0)^2+\frac{\ddot{f}_0}{6f_0}(t-t_0)^3
\label{eq:tau2}
\end{equation}
again the signal phase becomes that of a monochromatic signal which means that the spin-down shift has been removed. Note that in practice the spin-down correction is applied after barycentric corrections, then the time $t$ that appears in Eq.(\ref{eq:tau2}) is in fact the re-scaled time $\tau_1$ of Eq.(\ref{eq:baricorr}).

\section{\label{sec:targeted}Matched filter in the space of signal Fourier components}

Let us indicate the data at hand by 
\begin{equation}
x(t)=n(t)+h(t)
\end{equation}
where $n(t)$ is the noise and $h(t)$ is a \gw signal.
On this data the barycentric and spin-down corrections are applied as described in previous section. As a consequence, the signal is now monochromatic apart from an amplitude and phase sidereal modulation and is given by
\begin{equation}
h(t)=H_0(H_+A^++H_{\times}A^{\times})e^{\jmath \omega_0t+\gamma_0}
\label{eq:recsig}
\end{equation}
that is, it can be seen as the product of a {\it fast} periodic term, with frequency $f_0=\frac{\omega_0}{2\pi}$, and a {\it slow} term given by a linear combination of sines and cosines with argument $\Omega_{\oplus}$ and  $2\Omega_{\oplus}$, see Eqs.(\ref{eq:Ap},\ref{eq:Am}).  Then, the signal is completely described by its Fourier components at the 5 angular frequencies $\omega_0,~\omega_0\pm \Omega_{\oplus},~\omega_0\pm 2\Omega_{\oplus}$. This set of 5 complex numbers constitutes the signal {\it 5-vector}. Given a generic time series $g(t)$, the corresponding 5-vector is
\begin{equation}
\mathbf{G}=\int_T g(t)e^{-\jmath (\omega_0-\mathbf{k}\Omega_{\oplus})t}dt
\label{eq:def5vect}
\end{equation}
where $\mathbf{k}=[-2,-1,...,2]$ and $T$ is the observation time. In the following we indicate with $\mathbf{X}$ the data 5-vector and with $\mathbf{A}^+,~\mathbf{A}^{\times}$ the signal plus and cross 5-vectors, obtained by applying the definition of Eq.(\ref{eq:def5vect}) to Eqs.(\ref{eq:Ap},\ref{eq:Am}). These two last quantities depend only on known parameters and form the signal templates.   

The data 5-vector is
\begin{equation}
\mathbf{X}=H_0e^{\jmath \gamma_0}\left({H}_{+}\mathbf{A}^{+}+{H}_{\times}\mathbf{A}^{\times}\right)+\mathbf{N}
\end{equation}
where $\mathbf{N}$ is the 5-vector of noise alone.
Once the 5-vectors of data and of signal templates have been computed, the two complex quantities
\begin{equation}
\hat{H}_{+/\times}=\frac{\mathbf{X}\cdot \mathbf{A}^{+/\times}}{|\mathbf{A}^{+/\times}|^2}
\label{eq:hphc}
\end{equation}
are built, see \cite{ref:vela_vsr2}, \cite{ref:fivevect} for more details. They correspond to compute two matched filters between the data and the signal templates. Assuming the noise is Gaussian with mean value zero, it can be shown that these two quantities are estimators of the signal plus and cross amplitudes $H_0e^{\jmath \gamma_0}H_+,~H_0e^{\jmath \gamma_0}H_{\times}$. The estimators of Eq.(\ref{eq:hphc}) are used to build the detection statistic 
\begin{equation}
\mathcal{S}=|\mathbf{A}^{+}|^4|\hat{H}_{+}|^2+|\mathbf{A}^{\times}|^4|\hat{H}_{\times}|^2
\label{eq:detstat}
\end{equation}

\section{\label{sec:pvalue}Assessment of detection significance}

According to the {\it frequentist} paradigm, we can use the value of the detection statistic actually obtained in a given analysis, $\mathcal{S}^*$, to establish if our result is compatible with pure noise or not. This is done by computing the {\it p-value}, that is the probability that a value of the detection statistic equal or larger than $\mathcal{S}^*$ can be obtained analyzing noise only data,
\begin{equation}
p^*=P(\mathcal{S}\ge \mathcal{S}^*|h=0),
\label{eq:pvalue}
\end{equation}
and comparing it to a chosen threshold $p_{thr}$. In targeted searches of CW a typical choice for the threshold is $p_{thr}=0.01$ or less. If $p^*>p_{thr}$ we conclude our data are fully compatible with noise, otherwise we have a potentially {\it interesting} candidate, deserving a deeper study. Given the nature of CW signals, making a deeper study basically means analyzing longer and longer stretches of data, belonging to the same or to another detector, and computing the corresponding {\it p-values}. In case a real signal is present into the data we expect to find a smaller and smaller {\it p-value} until the detection can be claimed with high statistical confidence. At this point, signal unknown parameters, $H_0, ~\eta, ~\psi, ~\gamma_0$ can be estimated using proper combinations of the real and imaginary parts of the two complex amplitude estimators of Eq.(\ref{eq:hphc}), as described in \cite{ref:fivevect}. 

\begin{figure*}[!htbp]
\includegraphics[width=1.0\textwidth]{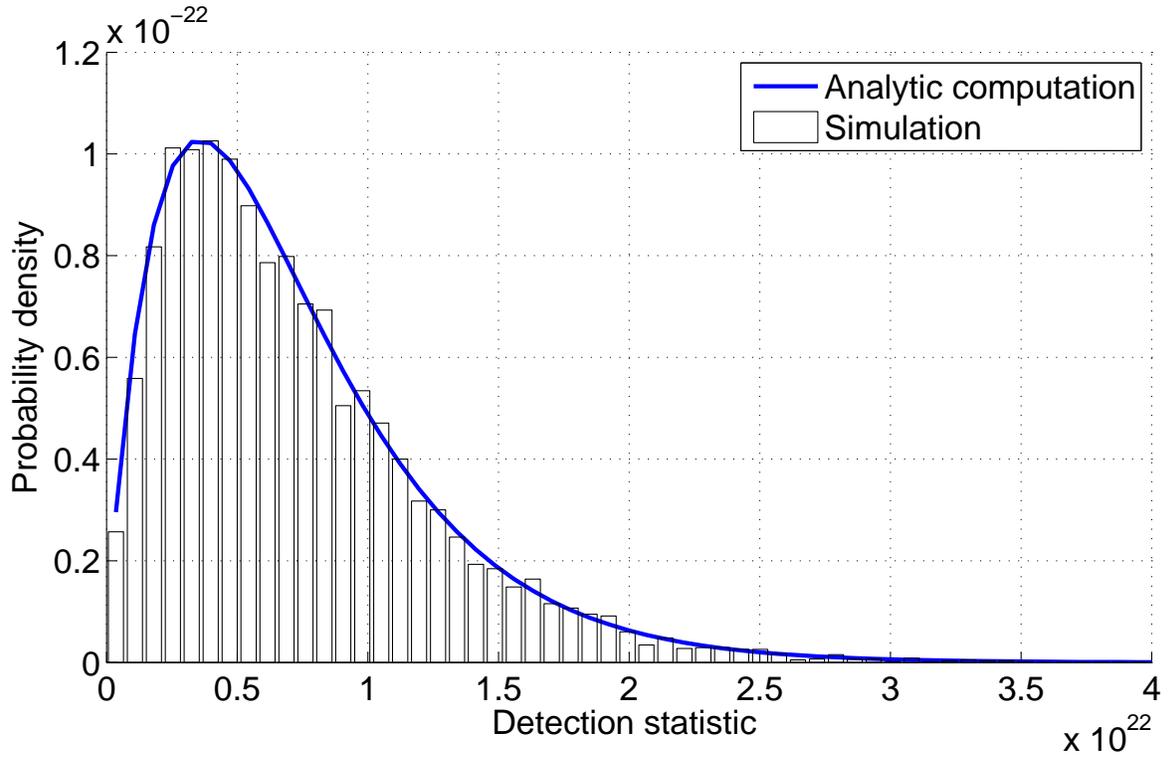}
\caption{Noise probability density of the detection statistic computed assuming noise is Gaussian with mean value zero and variance $\sigma^2=1$ and taking a total observation time $T=10^7$ seconds. The continuous line is obtained using the analytic formula given by Eq.(\ref{eq:pdfS}), the histogram is the result of a simulation.
\label{fig:noise_detstat_pdf}}
\end{figure*}

\begin{figure*}[!htbp]
\includegraphics[width=1.0\textwidth]{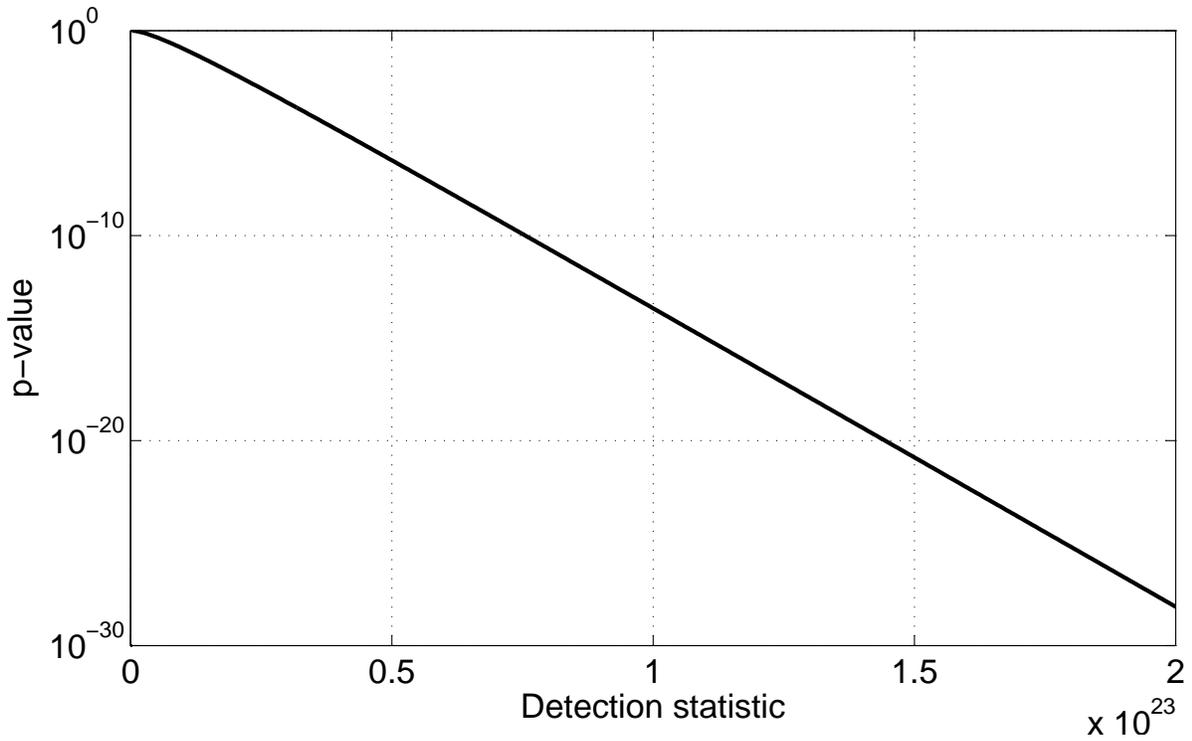}
\caption{Probability of having a value of the detection statistic larger than a value in the abscissa in case of noise only. It has been computed using the analytic formula of Eq.(\ref{eq:Pabove}), with the same choices of Fig.(\ref{fig:noise_detstat_pdf}). 
\label{fig:noise_pabove}}
\end{figure*}

In order to compute the {\it p-value}, defined by Eq.(\ref{eq:pvalue}), we need to know the probability distribution of the detection statistic for noise only. The theoretical distribution can be analytically derived under the assumption that the noise is Gaussian, with mean value zero and variance $\sigma^2$. From the definition of 5-vector, Eq.(\ref{eq:def5vect}), it follows that each component of the noise 5-vector is also distributed according to a gaussian with mean value zero and variance $\sigma^2_X=\sigma^2\cdot T$. It is easy to see that also the two complex amplitude estimators of Eq.(\ref{eq:hphc}) have a Gaussian distribution with zero mean value and variance
\begin{equation}
\sigma^2_{+/\times}=\frac{\sigma^2_X}{|\mathbf{A}^{+/\times}|^2}  
\label{eq:varest}
\end{equation}
The probability density function of the square modulus of the estimators is then exponential and given by
\begin{align}
f(x)=\frac{|\mathbf{A}^{+/\times}|^2}{\sigma^2_X}e^{-\frac{|\mathbf{A}^{+/\times}|^2}{\sigma^2_X} x}; ~~~~~x=|\hat{H}_{+/\times}|^2
\label{eq:ampestpdf}
\end{align}
From here we can derive the probability density for the detection statistic:
\begin{equation}
f(\mathcal{S})=\frac{e^{-\frac{\mathcal{S}}{\sigma^2_X|\mathbf{A}^{\times}|^2}}-e^{-\frac{\mathcal{S}}{\sigma^2_X|\mathbf{A}^{+}|^2}}}{\sigma^2_X\left(|\mathbf{A}^{\times}|^2-
|\mathbf{A}^{+}|^2\right)}
\label{eq:pdfS}
\end{equation}
In Fig.(\ref{fig:noise_detstat_pdf}) the detection statistic noise probability density given by Eq.(\ref{eq:pdfS}) is shown and compared to the result of a Monte Carlo simulation.
The probability of finding a value of the detection statistic above a given value $\mathcal{S}^*$, that is the {\it p-value}, is 
\begin{equation}
P(\mathcal{S}>\mathcal{S}^*)=\frac{|\mathbf{A}^{\times}|^2e^{-\frac{\mathcal{S}^*}{\sigma^2_X|\mathbf{A}^{\times}|^2}}-|\mathbf{A}^{+}|^2e^{-\frac{\mathcal{S}^*}{\sigma^2_X|\mathbf{A}^{+}|^2}}}{|\mathbf{A}^{\times}|^2-|\mathbf{A}^{+}|^2}
\label{eq:Pabove}
\end{equation}
and is plotted in Fig.(\ref{fig:noise_pabove}) under the same assumptions of Fig.(\ref{fig:noise_detstat_pdf}).
In practice, real data can show departure from gaussianity. The noise probability distribution can then be built from the data itself considering a range of {\it off-source} frequencies near but different from the one where the signal is supposed to be.

\section{\label{sec:motiv}Motivations for narrow-band searches}

Given the uncertainties on gravitational wave emission mechanisms and also the lack of a full detailed picture of the electro-magnetic emission geometry, it is not obvious at all that the gravitational-wave emission takes place at {\it exactly} twice the star measured pulse rate, or that such relation holds for observation times of months to years. For instance, if a neutron star is made of a crust and a core rotating at a slightly different rate, and if the gravitational-wave emission is dominated by an asymmetry in the core then a search targeted at 2$f_{rot}$ would assume a wrong signal frequency.   
Following the discussion in \cite{ref:crab_nb} we describe such situation by allowing the signal frequency to vary in the range:
\beq
f(t) \in \left[f_0(t)(1-|\delta|),~ f_0(t)(1+|\delta|)\right ] 
\label{eq:widef}
\eeq  
where $f_0(t)=f_0+\dot{f}_0t+\frac{\ddot{f}_0}{2}t^2$ is the signal frequency we would have if the gravitational-wave and electromagnetic signals were locked (as anticipated $f_0(t)=2f_{rot}(t)$ in the case of a non-axisymmetric neutron star rotating around one of its principal axis of inertia) and $\delta$ is a small positive or negative shift. The width of this range is $\Delta f=2f_0|\delta|$. If the star two spinning components are linked by some torque which tends to enforce corotation on a timescale $\tau_c$, then $\delta \sim \tau_c /\tau_{sd}$, where $\tau_{sd} \sim f_0/\dot{f}_0$ is the characteristic spin-down time. A relation of the form of Eq.(\ref{eq:widef}) also holds in the case the gravitational radiation is produced by free precession of a nearly bi-axial star \cite{ref:jones1}, in which case $\delta$ is of the order of $(I_{zz}-I_{xx})/I_{xx}$. In general, a value of $\delta$ of the order of, say, $10^{-4}$, corresponds to $\tau_c \sim 10^{-4}\tau_{sd}$ which, depending on the specific targeted pulsar can be of several months or years. This would be comparable or larger than the longest timescale observed in pulsar glitch recovery where a recoupling between the two component might occur. In terms of free precession, $|\delta|\sim 10^{-4}$ is on the high end of the range of deformations that neutron stars could be able to sustain \cite{ref:owen},\cite{ref:hask}. 
Concerning the spin-down range, at least in the case of the two-component model the existence of a torque that tends to enforce corotation implies that the the spin-down shift, $\dot{f}-\dot{f}_0$, is not independent on the frequency shift: if $f>f_0$ then $|\dot{f}|>|\dot{f}_0|$ and viceversa so that the frequency difference tends to decrease in time. As shown in Sec. \ref{sec:anagen}, the method we use to correct spin-down naturally goes in this direction and for each first order spin-down value $\dot{f}$ we explore a range $\Delta \dot{f}=2|\dot{f}_0\delta|$ around it. Similarly, for the second order spin-down we would explore a range $\Delta \ddot{f}=2|\ddot{f}_0\delta|$ around each allowed value, even if in practice we will see that typically this is not needed. 

\section{\label{sec:anagen}Search method description}

In a \nbs the source position is assumed to be known, while a range of values for the frequency $\Delta f$ and the spin-down terms $\Delta \dot{f}, \Delta \ddot{f}, ...$ is explored. The corresponding number of points in the source parameter space is then given by the product between the number of frequency bins $n_{freq}$ and the number of spin-down bins $n_{sd}=\prod_{i} n^{(i)}_{sd}$, where $n^{(i)}_{sd}$ is the number of spin-down values of order $i^{th}$ and only terms for which $n^{(i)}_{sd}\ge 1$ are considered. The width of the frequency bin is $\delta f=\frac{1}{T}$ then the number of frequency bins to be considered is
\begin{equation}
n_{freq}=\left[ \frac{\Delta f}{\delta f}\right ]=\left[ \Delta f \cdot T\right]\approx 6.3\cdot 10^5\left(\frac{\Delta f}{0.02~Hz} \right)\left(\frac{T}{1~yr} \right)
\label{eq:nfreq}
\end{equation}
where $\left[ ~\cdot{}~ \right]$ stands for the nearest integer. The bin width for spin-down of order $i^{th}$ is computed by imposing that an uncorrected amount of one bin produces a frequency variation over the observation time $T$ at most equal to half a frequency bin:
\begin{equation}
\frac{\delta f^{(i)}\cdot T^i}{i ! }=\frac{\delta f}{2}
\label{eq:sdbin}
\end{equation}
hence for the first and second spin-down order we find
\begin{eqnarray}  
\delta \dot f=\delta f^{(1)}=\frac{1}{2T^2}\\ 
\delta \ddot f=\delta f^{(2)}=\frac{1}{T^3}
\label{eq:sdbin2}
\end{eqnarray}
Consequently, the number of bins for first order spin-down is
\begin{align}
n^{(1)}_{sd}=\left[2\Delta \dot{f}\cdot T^2\right]\\ \nonumber
\approx 400\left(\frac{\dot{f}_0}{10^{-10}~Hz/s} \right)\left(\frac{\delta_0}{10^{-3}} \right)\left(\frac{T}{1~yr} \right)^2
\label{nsd}
\end{align} 
while for second order spin-down we have
\begin{align}
n^{(2)}_{sd}=\left[\Delta \ddot{f}\cdot T^3\right]\\ \nonumber
\approx 0.6\left(\frac{\ddot{f}_0}{10^{-20}~Hz/s^2} \right)\left(\frac{\delta_0}{10^{-3}} \right)\left(\frac{T}{1~yr} \right)^3
\label{eq:nsd2}
\end{align} 
For observation times of the order of the year and range of spin-down values typical of narrow-band searches, the corresponding number of bins is bigger than one only for the first order term, then $n_{sd}=n^{(1)}_{sd}$. In fact, the two previous equations would be correct if the explored range of spin-down values were independent of the frequency. As a matter of fact, as we will see below, this is not the case with the procedure we use to correct spin-down, for which the natural variable to consider is, rather than the spin-down, the ratio 
$\frac{\dot{f}}{2f}$.  

\begin{figure*}[!htbp]
\includegraphics[width=1.0\textwidth]{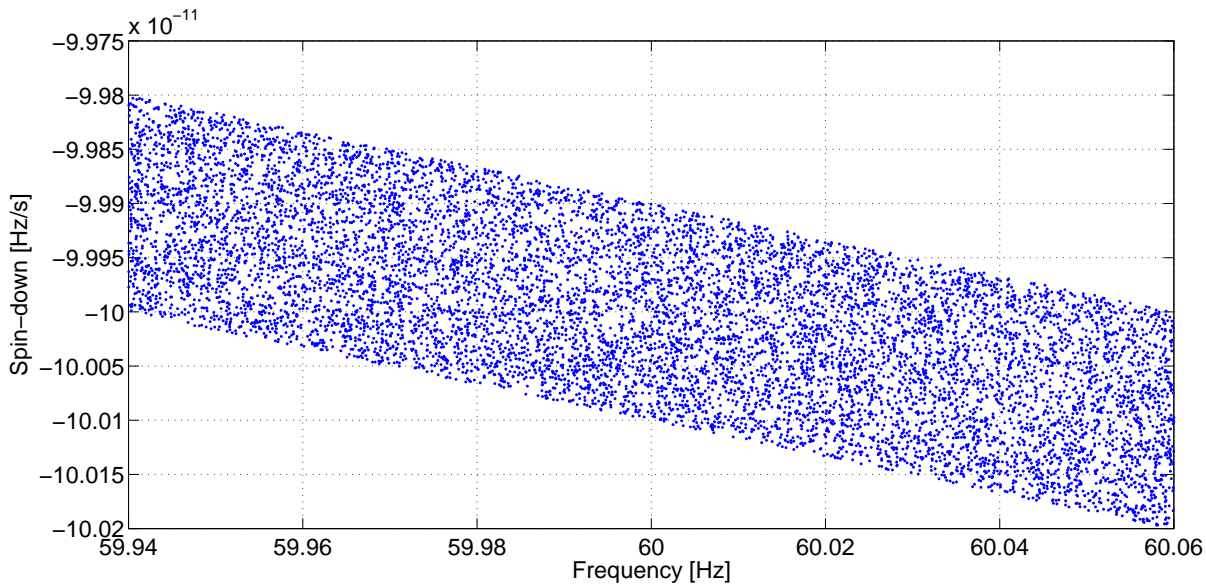}
\caption{Portion of the $f-\dot{f}$ plane covered in a narrow-band search with $f_0=60$ Hz, $\dot{f}_0=-10^{-10}$ Hz/s and $\delta=10^{-3}$. 
\label{fig:ffdot}}
\end{figure*}

In principle, according to Eq.(\ref{eq:fdopp}) for each frequency bin a barycentric correction should be applied. This is what has been done, e.g. in \cite{ref:crab_nb}. This 'brute force' approach becomes computationally heavier and heavier as the number of frequency bins increases.
By using the time-domain procedure described in Sec.\ref{sec:bary}, which as we have seen is independent on the frequency, barycentric corrections must be computed just once and hold for the whole frequency band. On the other hand, spin-down corrections are done using Eq.(\ref{eq:tau2}), which explicitly depends on the ratio $\frac{\dot{f}}{f}$. 
In fact we consider the quantity $\lambda=\frac{\dot{f}}{2f}$ as the independent variable, so that a grid is built on $(f, \lambda)$ rather than on $(f, \dot{f})$. In terms of $\lambda$ the time transformation that allows to correct the spin-down, given by Eq.(\ref{eq:tau2}), can be written as
\begin{equation}
\tau_2=t+\lambda\left(t-t_0\right)^2
\label{eq:tau2_lambda}
\end{equation}
The range of values for $\lambda$ is taken with width
\begin{equation}
\Delta \lambda=\frac{\Delta \dot{f}}{2f_{min}}
\label{eq:Dlambda}
\end{equation}
where $f_{min}=f_0(1-|\delta|)$ is the minimum analyzed frequency, and is centered around $\lambda_0=\frac{\dot{f}_0}{2f_0}$:
\be
\lambda \in [\lambda_{min}=\lambda_0-\frac{\Delta \lambda}{2}, \lambda_{max}=\lambda_0+\frac{\Delta \lambda}{2}]
\label{eq:lamra}
\ee
The corresponding {\it actual} search band for each value of $\dot{f}$, which varies in the range between $\dot{f}_0\frac{f_{min}}{f_0}$ and $\dot{f}_0\frac{f_{max}}{f_0}$, being 
$f_{max}=f_0(1+|\delta|)$, is then delimited by $\dot{f}_{min}=2f\lambda_{min}$ and $\dot{f}_{max}=2f\lambda_{max}$ with width equal to $2f(\lambda_{min}-\lambda_{max})$. This is equal to $\Delta \dot{f}$ at $f=f_{min}$ and increases for increasing value of $f$.
This is due to the conservative choice done in Eq.(\ref{eq:Dlambda}), where the minimum frequency $f_{min}$ has been taken at the denominator in such a way to maximize the range for $\lambda$. This means that the area of the explored parameter space is slightly larger than that we would have if the spin-down range was independent of the frequency.
In Fig.(\ref{fig:ffdot}) the portion of the $f-\dot{f}$ plane covered in a narrow-band search with $f_0=60$ Hz, $\dot{f}_0=-10^{-10}$ Hz/s and $\delta=10^{-3}$ is shown. Each point has been obtained choosing 10000 random values of the frequency between $f_{min}$ and $f_{max}$ and of $\lambda$ according to Eq.(\ref{eq:lamra}) and computing the corresponding $\dot{f}$ values.  

The interval in $\lambda$ is divided in a number of bins which width is, conservatively, taken as
\begin{equation}
\delta \lambda= \frac{\delta \dot{f}}{2f_{max}}
\label{eq:dlambda}
\end{equation}
The number of $\lambda$ values to be taken into account is then 
\begin{equation}
n_{\lambda}=\left[ \frac{\Delta \lambda}{\delta \lambda} \right]=\left[ n^{(1)}_{sd}\frac{1+|\delta|}{1-|\delta|}\right]
\label{eq:nlambda}
\end{equation}  
where $n^{(1)}_{sd}$ is the ``canonical'' number of spin-down values of the first order, given by Eq.(\ref{nsd}).
For $|\delta|\ll 1$ (or, equivalently, $\frac{f_{max}}{f_{min}}\approx 1$) which is typical of a standard narrow-band search, we have $n_{\lambda}\simeq n^{(1)}_{sd}$. In fact, we must stress that the time-domain correction method described in this section can be applied whatever range of frequencies is analyzed. In particular, it can be used also for the so-called {\it directed} searches of neutron stars which position is known but which frequency, and spin-down, are completely unknown because no electro-magnetic pulsation is observed, see e.g. \cite{ref:wette}. For such kind of analysis the number of different values of $\lambda$ to be considered will be larger than the ``canonical'' number of spin-down values, as $\frac{f_{max}}{f_{min}}\gg 1$.   

Once barycentric corrections have been done and, for each value of $\lambda$, also the spin-down has been corrected, the detection statistic, see Eq.(\ref{eq:detstat}), is computed for each frequency in the considered range $\Delta f$. Then we end with $n_{freq}\cdot n_{\lambda}$ values of the detection statistic. The maximum of this set of numbers, let us call it $\mathcal{S}_{max}$, corresponds to the most significant candidate of the analysis (also called 'loudest event') and is used to compute the corresponding {\it p-value}. If it results to be compatible with noise, then an upper limit is established. 

In fact, exploring a large number of points in the source parameter space has an impact on the statistical significance of the analysis results, because the probability that noise alone produces a value of the detection statistic larger than the value actually found in the analysis is clearly larger than in the case of search in a single point of the parameter space. This is the well-known {\it look-elsewhere effect} (or {\it trial factor}), see e.g. \cite{ref:wette2}. In practice, we want to assess the statistical significance of a given analysis result by using the same procedure described in Sec.\ref{sec:pvalue}, that is by considering the single-trial noise probability density of Eq.(\ref{eq:pdfS}). This can be done provided we use a suitable threshold for discriminating between "interesting" and "non-interesting" candidates, different from that we would use for a {\it targeted} search and determined in the following way.  Let us indicate with $p_{thr}$ the overall significance threshold for a \nbs, e.g. $p_{thr}=0.01$. Let us indicate with $p_0$ the corresponding significance level computed over the single-trial noise distribution and assume that the searches in each point of the parameter space are independent. The probability $p_{above}$ that at least one of the results is significant, i.e. that it gives a value of the detection statistic above $\mathcal{S}_{p_0}$, is equal to 1 minus the probability that none of them are significant. Since it is assumed that they are independent, the probability that all of them are not significant is the product of the probabilities that each of them are not significant, that is $\left(1-p_0\right)^N$, where $N=n_{freq}\cdot n_{\lambda}$ is the number of points in the parameter space. Then, we have $p_{above}=1-\left(1-p_0\right)^N$, By imposing $p_{above}=p_{thr}$ and solving for $p_0$ we obtain
\begin{equation}
p_0=1-\left(1-p_{thr}\right)^{\frac{1}{N}}
\label{p0}
\end{equation}
For $p_{thr}\ll 1$ and $N$ large, like in our case, we have
\begin{equation}
p_0\simeq \frac{p_{thr}}{N}
\label{p0approx}
\end{equation} 
Extending Eq.(\ref{eq:pvalue}) we indicate with 
\begin{equation}
p_{loudest}=P\left(\mathcal{S}>\mathcal{S}_{max}|h=0\right)
\label{ploud}
\end{equation}
the {\it p-value} corresponding to the loudest candidate of the analysis. A potentially interesting candidate is such that $p_{loudest}<p_0$.

\section{\label{sec:sensi}Search sensitivity}

\begin{figure*}[!htbp]
\includegraphics[width=1.0\textwidth]{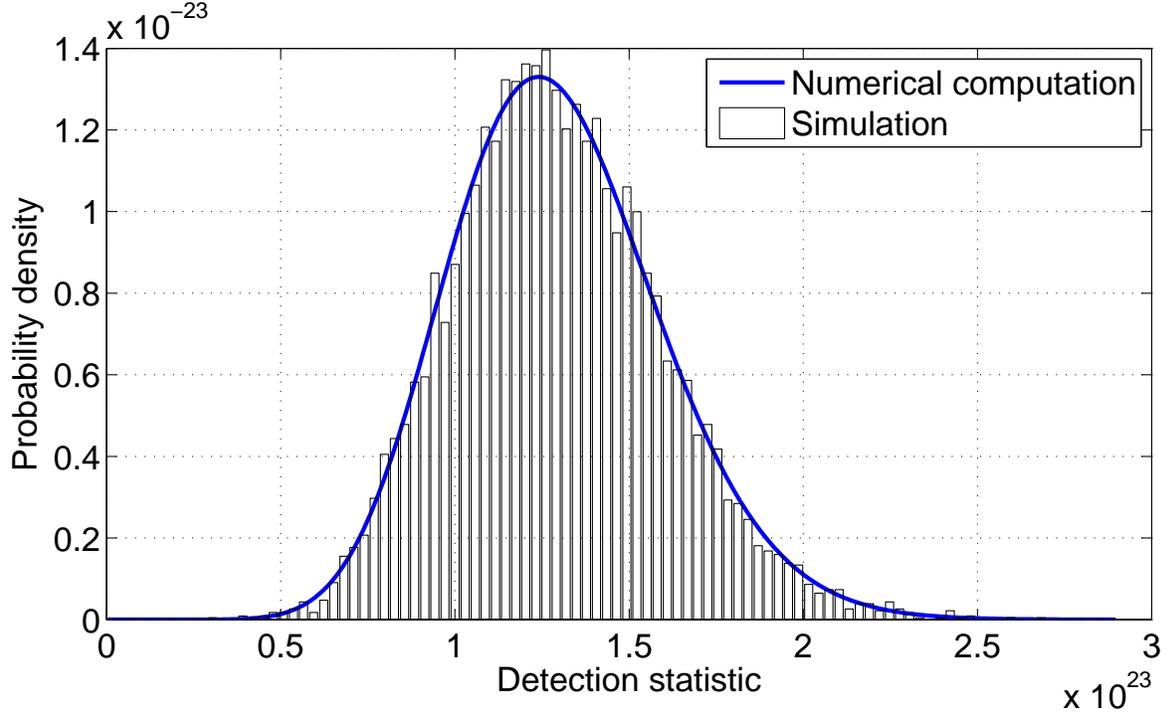}
\caption{Probability distribution of the detection statistic considering a signal of amplitude $H_0=0.038$ summed to gaussian noise with $\sigma =1$ over $10^7$ seconds. The continuous line is the result of the numerical integration of Eq.(\ref{eq:pdf_S_sig}), while the histogram has been obtained with a Monte Carlo simulation. 
\label{fig:distr_sig_noise}}
\end{figure*}

\begin{figure*}[!htbp]
\includegraphics[width=1.0\textwidth]{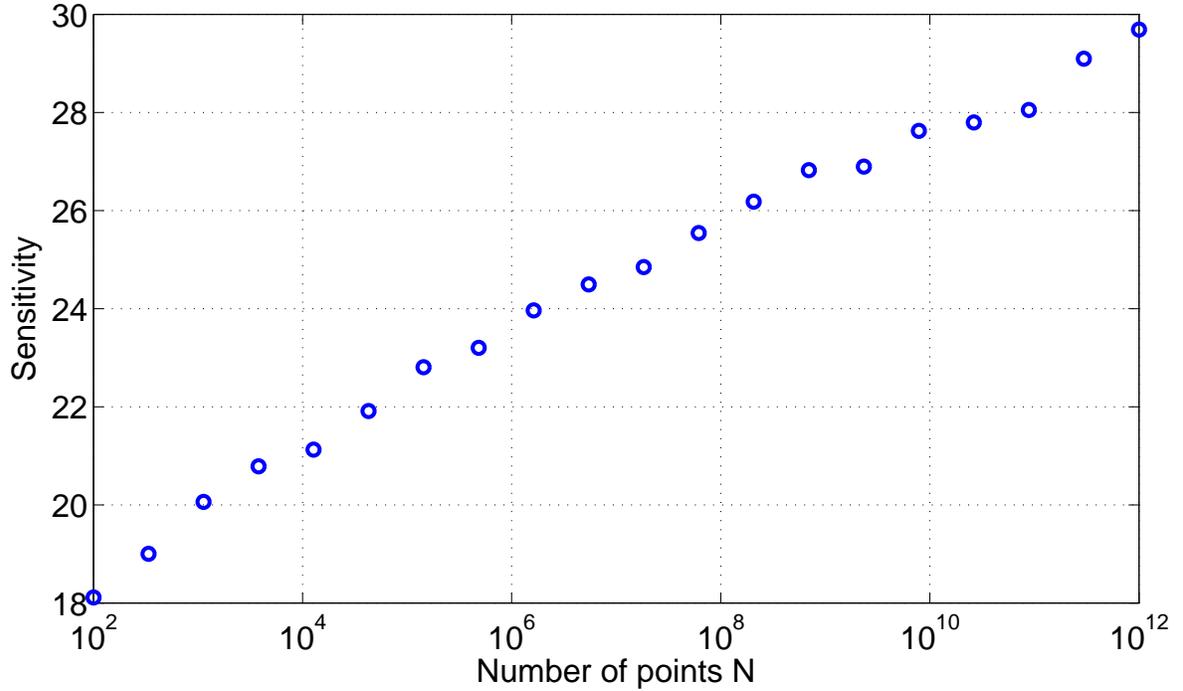}
\caption{Sensitivity of the narrow-band search in units of $\sqrt{\frac{S_n}{T}}$ as a function of the number of points in the parameter space, with the choice $p_{thr}=0.01$ and $p_{det}=0.95$. An uncertainty of $\sim 5\%$ is associated to the values, due to the finite size of the simulation.
\label{fig:sensitivity_narrow}}
\end{figure*}

The search sensitivity for CW signals is defined as the minimum strain amplitude, $h_{min}$, detectable with a given detection probability $p_{det}$ and at a fixed {\it p-value} $p_{thr}$. To compute the theoretical sensitivity we need first to know the probability density distribution of the detection statistic in case a signal of amplitude $H_0$ is present into the noise. This can be done straightforwardly starting from the distribution of the real and imaginary part of the amplitude estimators $\hat{H}_+,~\hat{H}_{\times}$ given by Eq.(\ref{eq:hphc}). 
The resulting distribution for the square modulus of the estimators, which reduces to Eq.(\ref{eq:ampestpdf}) if $H_0=0$, is given by
\begin{equation}
f(x;H_0)=\frac{k}{2}e^{-\frac{1}{2}\left(kx+\beta\right)}I_0\left(\sqrt{k\beta x}\right)
\label{eq:pdf_sig}
\end{equation}
where, as in Eq.(\ref{eq:ampestpdf}), $x=|\hat{H}_{+/\times}|^2$ while $k=2\frac{|\mathbf{A}^{+/\times}|^2}{\sigma^2_X}$, $\beta=2\frac{H_0^2|e^{\jmath \Phi_0}H_{+/\times}\mathbf{A}^{+/\times}|^2}{\sigma^2_X}$ and $I_0$ is the modified Bessel function of the first kind of order zero. Eq.(\ref{eq:pdf_sig}), which is derived in Appendix \ref{app:app1}, describes, apart from the factor $k$, a $\chi ^2$ distribution with two degrees of freedom. From this equation is immediate to write down the probability distribution for the variables $y_+=|\mathbf{A}^{+}|^4 |\hat{H}_{+}|^2$ and $y_{\times}=|\mathbf{A}^{\times}|^4 |\hat{H}_{\times}|^2$:
\begin{align}
f(y_+) &=\frac{e^{-\left(\frac{y_+}{\mu_+}+\frac{\beta_+}{2} \right)}}{\mu_+}I_0\left(\sqrt{2y_+\frac{\beta_+}{\mu_+}}\right) \notag \\
\mu_+ &=\sigma^2_X|\mathbf{A}^{+}|^2 \\
\beta_+ &=2\frac{H_0^2|e^{\jmath \Phi_0}H_{+}\mathbf{A}^{+}|^2}{\sigma^2_X} \notag 
\label{eq:pdf_yplus}
\end{align}
\begin{align}
f(y_{\times}) &=\frac{e^{-\left(\frac{y_{\times}}{\mu_{\times}}+\frac{\beta_{\times}}{2} \right)}}{\mu_{\times}}I_0\left(\sqrt{2y_{\times}\frac{\beta_{\times}}{\mu_{\times}}}\right) \notag \\
\mu_{\times} &=\sigma^2_X|\mathbf{A}^{{\times}}|^2 \\
\beta_{\times} &=2\frac{H_0^2|e^{\jmath \Phi_0}H_{\times}\mathbf{A}^{\times}|^2}{\sigma^2_X} \notag 
\label{eq:pdf_ycross}
\end{align}
Now, the distribution of the detection statistic $\mathcal{S}=y_+ + y_{\times}$ is given by the convolution of the two distributions $f(y_+)$ and $f(y_{\times})$:
\begin{widetext}
\begin{equation}
f(\mathcal{S})=\frac{e^-{\frac{\beta_+ + \beta_{\times}}{2}}}{\mu_+ \mu_{\times}}e^{-\frac{\mathcal{S}}{\mu_+}}\int_0^{\mathcal{S}}e^{-\left(\frac{1}{\mu_{\times}}-\frac{1}{\mu_+} \right)y_{\times}}I_0\left(\sqrt{2(\mathcal{S}-y_{\times})\frac{\beta_+}{\mu_+}}\right)I_0\left(\sqrt{2y_{\times}\frac{\beta_{\times}}{\mu_{\times}}}\right)dy_{\times}
\label{eq:pdf_S_sig}
\end{equation}
\end{widetext}
In Fig.(\ref{fig:distr_sig_noise}) the probability distribution of Eq.(\ref{eq:pdf_S_sig}) is plotted considering, as an example, a signal of amplitude $H_0=0.038$ summed to gaussian noise with $\sigma =1$ over $10^7$ seconds.
The integral of Eq.(\ref{eq:pdf_S_sig}) can be evaluated numerically for given signal parameters. 
Alternatively, to take into account a possible departure of the noise from gaussianity, the probability distribution of $\mathcal{S}$ in presence of a signal can be built through a Monte Carlo simulation in which several signals are generated in software and summed to the data and the resulting value of the detection statistic computed. Whatever approach is used, once this distribution is known the sensitivity is estimated in two steps: from the noise-only distribution of the detection statistic we first compute the value $\mathcal{S}_{thr}$ corresponding to the chosen {\it p-value}, $p_{thr}$; then we determine the signal amplitude $h_{min}$ such that $P(\mathcal{S}>\mathcal{S}_{thr}|h_{min})=p_{det}$. We can express the sensitivity explicitly showing the dependency on the detector noise spectrum at the signal frequency, $S_n$, and the observation time $T$:
\begin{equation}
h_{min}=\mathcal{C}\sqrt{\frac{S_n}{T}}
\label{eq:hmin}
\end{equation}
where the coefficient, $\mathcal{C}$ depends on the number of points in the parameter space, $N$, and on the chosen values of $p_{thr}$ and $p_{det}$. The coefficient $\mathcal{C}$ is plotted in Fig.(\ref{fig:sensitivity_narrow}) as a function of $N$ for $p_{thr}=0.01$ and $p_{det}=0.95$.   
From the figure we see that for the considered range of values of $N$ the sensitivity of the narrow-band search is weekly dependent on $N$ and is $\sim$2-3 times worse than the sensitivity of a {\it targeted} search, which is characterized by $\mathcal{C}\approx 11$, the exact value depending on the specific analysis method. This is expected as a consequence of the volume of the explored parameter space, as discussed in Sec. \ref{sec:anagen}. 

Two obvious targets for a narrow-band search would be the Crab (J0534+2200) and Vela (J0835-4510) pulsars. For these two pulsars the {\it spin-down limit}, see Eqs.(\ref{eq:h0sd},\ref{eq:eps_sd}), has been beaten setting experimental upper limits in {\it targeted} searches with data of Virgo and LIGO detectors \cite{ref:abbott2010}, \cite{ref:vela_vsr2}, \cite{ref:aasi2013} that allow, in the latest analyses, to constrain the fraction of rotational energy lost through the emission of gravitational waves to, respectively, about 1$\%$ and 10$\%$. As we have seen, the sensitivity of a narrow-band search is a factor of 2-3 worse (depending on the extension of the explored parameter space) with respect to that of a {\it targeted} one. It is, however, interesting to see if for those specific sources the estimated sensitivity is still below the {\it spin-down limit} considering the most recent detector data. This would be a strong argument in favor of actually making such analysis. 
Let us consider Virgo VSR4 data, which are known to have a good low frequency sensitivity. Let us assume to make a search over a frequency band of $\Delta f=0.02$ Hz around the central gravitational wave frequency (which is $59.46$ Hz for Crab and $22.38$ Hz for Vela). Using the relations given in Sec. \ref{sec:anagen} we find for the Crab (and Vela) pulsars that $\delta=1.68\cdot 10^{-4}$ ($\delta=4.47\cdot 10^{-4}$), a number of frequency bins of $1.6\cdot 10^5$ ($1.6\cdot 10^5$), a width for  the first order spin-down range $\Delta \dot{f}=2.49\cdot 10^{-13}$ Hz/s ($2.81\cdot 10^{-14}$ Hz/s), corresponding to 33 (3) first order spin-down bins, while no further values of the second order spin-down must be considered. Overall, the total number of points in the parameter space is $5.28\cdot 10^6$ ($4.80\cdot 10^5$). By considering a typical Virgo VSR4 sensitivity curve and the run duration ($T_{obs}\approx 90$ days) we find a sensitivity at 1$\%$ {\it p-value} and 95$\%$ detection probability of $h_{min}\approx 7\cdot 10^{-25}$ for Crab and $h_{min}\approx 3\cdot 10^{-24}$ for Vela. The latter estimation is comparable to the Vela {\it spin-down limit}, while for Crab the value is about a factor of 2 below it, and then we can expect to improve with respect to results of the analysis of previous LIGO S5 data \cite{ref:crab_nb}. These estimations clearly indicates the relevance of this analysis on VSR4 data. An even better sensitivity would be achieved for Crab by including in the analysis also LIGO S6 data, while no improvement is expected in the case of Vela due to the S6 poor sensitivity at the corresponding frequency.        
    
\section{\label{sec:test}Method validation}
\begin{figure*}[!htbp]
\includegraphics[width=1.0\textwidth]{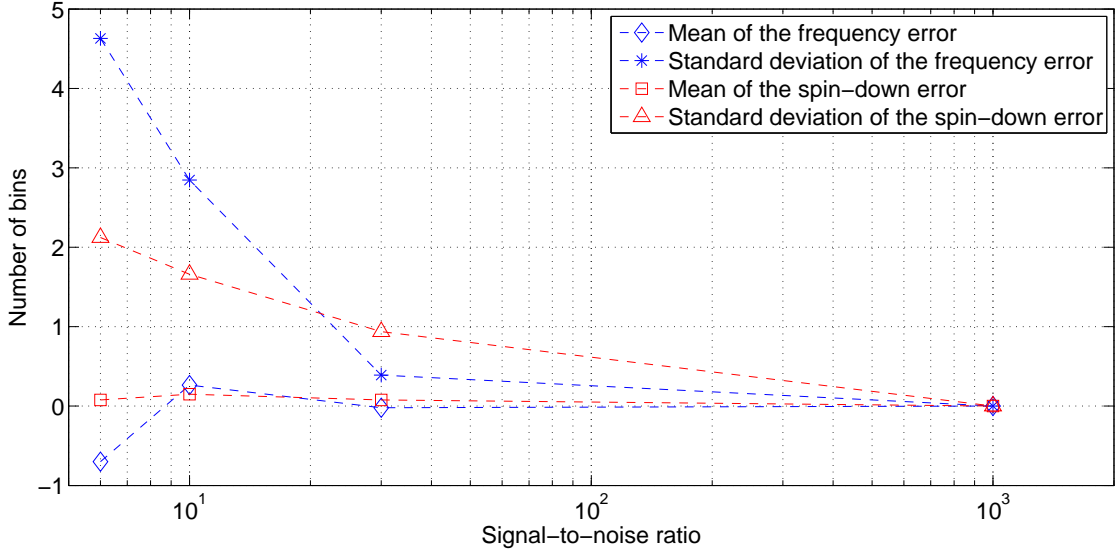}
\caption{Mean and standard deviation of the frequency and spin-down error as a function of the output signal-to-noise ratio. The error is expressed as a number of bins. For each value of the signal-to-noise ratio 30 signals with a given frequency and spin-down, with random position and polarization parameters, and lasting for $10^7$ seconds, have been generated and added to Gaussian noise. Then, a search has been done over $10^4$ frequency bins and 61 values of $\lambda$ and the frequency and spin-down values corresponding to the loudest event have been selected and compared to the injected values.
\label{fig:narrowband_errors}}
\end{figure*}

We have tested the analysis method through software injections into simulated Gaussian data. The barycentric and spin-down correction routines has been already verified, using both software and hardware injections, when used for {\it targeted} searches \cite{ref:vela_vsr2}. For narrow-band searches the main goal of the test has been to check if the method is able to recover the  frequency and spin-down of injected signals when a search over a range of parameters is done and the loudest event is selected. To do this for a given signal amplitude, frequency and spin-down, several signals have been generated (assuming, for computational reasons, that the Doppler effect correction has been already applied), corresponding to sources with random position and polarization parameters, and added to Gaussian noise. Then, the data have been analyzed and the frequency and spin-down of the loudest event have been estimated and compared to the injected values. This procedure has been repeated for various signal-to-noise ratios. The searched frequency band in the test was of 1 mHz corresponding to $10^4$ frequency bin, while the number of bins in the parameter $\lambda$ was 61. In Fig.(\ref{fig:narrowband_errors}) we report the mean and standard deviation of the error in frequency and spin-down estimation, expressed in number of bins, as a function of the output signal-to-noise ratio. Four different values of signal-to-noise ratio have been considered: a small value of 6, a moderate value of 10, a large value of 30, while the last value basically corresponds to the situation in which only the signal is present in the data. In this last case we see from the plot that frequency and spin-down are always perfectly recovered. For smaller and smaller values of the signal-to-noise ratio the errors increase as expected, but the distribution of results is still nearly centered at the correct parameter values.   

\section{\label{sec:concl}Conclusions}
In this paper we have described a coherent analysis method to perform narrow-band searches of continuous gravitational wave signals from known pulsars. Such kind of search allows us to take into account a possible small mismatch between the gravitational-wave signal frequency and two times the pulsar electromagnetic pulse rate. This difference could be due to various mechanisms and it is therefore important to have in place an analysis procedure robust against deviations from the standard assumptions of CW {\it targeted} searches, in which a strict correlations between the gravitational-wave signal frequency and the star rotation rate is assumed. 

 The use of an efficient time domain procedure to make barycentric corrections allows a large computational gain with respect to the standard 'brute force' approaches, like that used in \cite{ref:crab_nb}. Roughly speaking, given a search over $n_{freq}$ frequency bins the use of the time domain barycentric correction, that holds for all the frequencies, gives a computational gain of a factor $n_{freq}$. On the other hand, as discussed in Sec. \ref{sec:anagen}, for each considered spin-down value the detection statistic is computed for all the frequency bins. While the computation of a single value of the detection statistic takes a negligible time, its computation $n_{freq}$ times for each spin-down value becomes computationally relevant and partially reduces the gain due to the time domain barycentric corrections. Note that, on the contrary, if a 'brute force' method is used the computing time is largely dominated by the barycentric corrections. By making tests using 3 months of simulated data sets we have measured that in a search over $n_{freq}\approx 5\cdot 10^5$ frequency bins and of the order of 30 spin-down values the computational gain with respect to the 'brute force' approach is of the order of $10^4$. 

We have estimated the expected sensitivity for a narrow-band search, 0.02 Hz wide, around Crab and Vela pulsars using a typical Virgo VSR4 noise curve obtaining values which are comparable to the {\it spin-down limit} for Vela and about two times below for Crab. Interestingly, the value for the Crab is also about a factor of two below with respect to past published upper limits  \cite{ref:crab_nb} while the portion of the explored parameter space is a factor 1.7 larger. 
This estimations suggest that a narrow-band search of Virgo VSR4 data around Crab and Vela pulsars is worthwhile and will be accomplished in the near future.

Moreover, several pulsars are expected to be potentially interesting for advanced gravitational-wave detectors, which will start taking data in 2015-2016, and will be natural targets for a narrow-band search. For instance, when advanced detectors will reach their final configuration \cite{ref:obscen} we can estimate a narrow-band search sensitivity for Crab and Vela of, respectively, $h_{min}\approx  5\cdot 10^{-26}$ and $h_{min}\approx  2\cdot 10^{-25}$. These values are comparable, or even bit below, the expected signal amplitude emitted by a maximally strained neutron star with standard equation of state.   


\appendix

\section{Probability distribution of  the signal amplitude estimators}
\label{app:app1}
Let us consider one of the amplitude estimators, e.g. $\hat{H}_{+}=H_0e^{\jmath \gamma_0}H_+$, when the data is just gaussian noise with zero mean and variance $\sigma^2$. Its real and imaginary parts, that we call $\hat{H}_{R}$ and $\hat{H}_{I}$ respectively, are distributed 
according to a normal distribution with zero mean and variance $\frac{\sigma_+^2}{2}$, where $\sigma_+^2$ is given by Eq.(\ref{eq:varest}). Let us define two new variables, $\beta_R=\frac{\hat{H}_{R}}{\sigma_+/\sqrt{2}}$ and $\beta_I=\frac{\hat{H}_{I}}{\sigma_+/\sqrt{2}}$, which are then distributed according to a normal distribution with zero mean and variance one. Then it follows that the sum of their squares follows, in case of noise only, a $\chi ^2$ distribution with two degrees of freedom, i.e. an exponential. If a signal is present into the data the distribution of $\beta=\beta_1^2+\beta_2^2$ is a non-central $\chi ^2$ with two degrees of freedom: 
\begin{equation}
p(\beta;\lambda)=e^{-\left(\beta+\frac{\lambda}{2}\right)}I_0\left(\sqrt{\beta \lambda}\right)
\end{equation}
with non-centrality parameter 
\begin{equation}
\lambda=\left(E[\beta_1]\right)^2+\left(E[\beta_2]\right)^2 
\end{equation}
A straightforward calculation shows that 
\begin{equation}
\lambda=\frac{2H_0^2|e^{\jmath \gamma_0}H_+\mathbf{A}^{+}|^2}{\sigma_X^2}
\end{equation}
Hence, the distribution of $|\hat{H}_{+}|^2=\beta\frac{\sigma_+^2}{2}$ is
\begin{equation}
p(|\hat{H}_{+}|^2;\lambda)=\frac{1}{\sigma_+^2}e^{-\frac{1}{2}\left(\frac{2|\hat{H}_{+}|^2}{\sigma_+^2}+\lambda \right)}I_0\left(\sqrt{2\frac{|\hat{H}_{+}|^2}{\sigma_+^2}\lambda}\right)
\end{equation}
which, by introducing the factor $k=2\frac{|\mathbf{A}^{+/\times}|^2}{\sigma^2_X}$, becomes
\begin{equation}
p(|\hat{H}_{+}|^2;\lambda)=\frac{k}{2} e^{-\frac{1}{2}\left(k|\hat{H}_{+}|^2+\lambda \right)}I_0\left(\sqrt{k|\hat{H}_{+}|^2\lambda}\right)
\end{equation}
that is Eq.(\ref{eq:pdf_sig}). An equivalent expression can be obviously derived for $\hat{H}_{\times}$.

\begin{acknowledgments}
We want to thank the anonymous referees for the constructive comments that allowed us to improve the paper.
\end{acknowledgments}
 
\bibliography{narrow_band}

\end{document}